\documentclass[review]{elsarticle}

\usepackage{lineno,hyperref}
\usepackage{amsfonts} 
\usepackage{amsmath}
\usepackage{amssymb}
\usepackage{siunitx}
\usepackage[dvipsnames]{xcolor}
\definecolor{gr}{rgb}{0.0, 0.0, 0.0}

\journal{Journal of \LaTeX\ Templates}

\begin{document}

\begin{frontmatter}

\title{Computer-aided shape features extraction and regression models for predicting the ascending aortic aneurysm growth rate}

\author[UTV_address,Ansys_address]{Leonardo Geronzi \corref{mycorrespondingauthor}}
\author[UTV_address,Ansys_address]{Antonio Martinez}
\author[Ansys_address]{Michel Rochette}
\author[Ansys_address,INSA_address]{Kexin Yan}
\author[INSA_address]{Aline Bel-Brunon}
\author[Rennes_address]{Pascal Haigron}
\author[Rennes_address]{Pierre Escrig}
\author[Rennes_address]{Jacques Tomasi}
\author[Rennes_address]{Morgan Daniel}
\author[Dijon_address,Dijon2_address]{Alain Lalande}
\author[Dijon_address,Dijon2_address]{Siyu Lin}
\author[Dijon_address,Dijon2_address]{Diana Marcela Marin-Castrillon}
\author[Dijon_hospital_address]{Olivier Bouchot}
\author[Toulouse_address]{Jean Porterie}
\author[UTV_address]{Pier Paolo Valentini}
\author[UTV_address]{Marco Evangelos Biancolini}
\address[UTV_address]{University of Rome Tor Vergata, Department of Enterprise Engineering ``Mario Lucertini”, Rome, Italy}
\address[Ansys_address]{Ansys France, Villeurbanne, France}
\address[INSA_address]{University of Lyon, INSA Lyon, CNRS, LaMCoS, UMR5259, 69621 Villeurbanne, France}
\address[Rennes_address]{University of Rennes, CHU Rennes, Inserm, LTSI – UMR 1099, F-35000, Rennes, France}
\address[Dijon_address]{ICMUB Laboratory, CNRS 6302, University of Burgundy, 21078 Dijon, France}
\address[Dijon2_address]{Medical Imaging Department, University Hospital of Dijon, Dijon, France}
\address[Dijon_hospital_address]{Department of Cardio-Vascular and Thoracic Surgery, University Hospital of Dijon, Dijon, France}
\address[Toulouse_address]{Cardiac Surgery Department, Rangueil University Hospital, Toulouse, France}

\cortext[mycorrespondingauthor]{Corresponding author. Email address: leonardo.geronzi@uniroma2.it}

\begin{abstract}
Objective: ascending aortic aneurysm growth prediction is still challenging in clinics. In this study, we evaluate and compare the ability of local and global shape features to predict ascending aortic aneurysm growth.

Material and methods: 70 patients with aneurysm, for which two 3D acquisitions were available, are included. Following segmentation, three local shape features are computed: (1) the ratio between maximum diameter and length of the ascending aorta centerline, (2) the ratio between the length of external and internal lines on the ascending aorta and (3) the tortuosity of the ascending tract. By exploiting longitudinal data, the aneurysm growth rate is derived. Using radial basis function mesh morphing, iso-topological surface meshes are created. Statistical shape analysis is performed through unsupervised principal component analysis (PCA) and supervised partial least squares (PLS). Two types of global shape features are identified: three PCA-derived and three PLS-based shape modes.
Three regression models are set for growth prediction: two based on gaussian support vector machine using local and PCA-derived global shape features; the third is a PLS linear regression model based on the related global shape features. The prediction results are assessed and the aortic shapes most prone to growth are identified.
 
Results: the prediction root mean square error from leave-one-out cross-validation is: 0.112 mm/month, 0.083 mm/month and 0.066 mm/month for local, PCA-based and PLS-derived shape features, respectively. Aneurysms close to the root with a large initial diameter report faster growth.

Conclusion: global shape features might provide an important contribution for predicting the aneurysm growth.

\end{abstract}

\begin{keyword}
\texttt{Ascending aortic aneurysm, growth prediction, shape features, regression}
\end{keyword}

\end{frontmatter}

\section{Introduction}

The ascending aortic aneurysm (AsAA) represents a dangerous and potentially life-threatening condition consisting in a permanent dilatation of the aortic wall \cite{salameh2018thoracic}. If left untreated, the aneurysm can continue to grow and potentially burst, leading to severe bleeding and even death \cite{isselbacher2005thoracic}. 
The risk associated with AsAA is currently mainly determined by assessing the maximum diameter of the aorta from imaging techniques \cite{anfinogenova2022existing}. According to the medical guidelines, when the AsAA reaches a diameter of 50 mm, surgical repair should be performed \cite{elefteriades2022ascending}. To date, however, the use of the diameter alone as a \textcolor{gr}{predictor} of complications has proven insufficient \cite{papakonstantinou2021elective} and an accurate assessment of the patient-specific risk of rupture is still missing. Thus, there is a strong  need for new risk evaluation strategies that take into account \textcolor{gr}{additional} clinical parameters and biomarkers for the AsAA \cite{strimbu2010biomarkers}. Anatomy seems to play an important role in both diagnosis and therapy of the pathology: shape alterations, in fact, often cause functional impairments which in turn further accentuate anatomical abnormalities \cite{saliba2015ascending}.

The necessity of conducting longitudinal studies to follow the progression of cardiac diseases has already been discussed in literature \cite{sophocleous2022feasibility}. In recent years, a strong contribution to the development of prediction methods has been provided by machine learning techniques \cite{canchi2015review,ostberg2022machine}.

In this field, most research focused on the abdominal aorta \cite{groeneveld2018systematic,akkoyun2020predicting}. \textcolor{gr}{In predicting the abdominal aortic aneurysm growth, automatic methods to extract the diameter from three-dimensional imaging such as the maximally inscribed sphere method have already been developed \cite{gharahi2015growth}.}
The importance of considering the vessel shape for predicting the aneurysm evolution has been discussed in literature \cite{joly2020cohort,raut2013role}. Shum et al.  \cite{shum2011quantitative} described a set of classification methods based on local geometric features and wall thickness to identify patients at risk of abdominal aortic aneurysm rupture. Zhang et al. \cite{zhang2019patient} proposed a predictive model to estimate the abdominal aortic aneurysm progression based on Growth and Remodelling (G$\&$R) approaches and able to quantify the uncertainty related to the growth prediction. \textcolor{gr}{Jiang et al. \cite{jiang2020deep} used deep learning methods and G$\&$R techniques to predict the aneurysm diameter assessed by maximally inscribed sphere methods.} Do et al. \cite{do2018prediction} developed a Dynamical Gaussian Process Implicit Surface approach to predict the evolution of abdominal aortic aneurysms.

Regarding the thoracic aorta, Sophocleous et al. \cite{sophocleous2019aortic} investigated the associations between morphological features and cardiovascular function in patients affected by aortic coarctation and with bicuspid aortic valve. 
The importance of  predicting post-operative risks related to aortic surgery was described by Ren et al. \cite{ren2022performance}. In \cite{lee2013surface}, a classifier based on the curvature of the thoracic aorta was proposed to evaluate the risk of rupture.

The 3D information \textcolor{gr}{related to the global shape} derived from the segmented images can be exploited using statistical shape analysis (SSA)  \cite{heimann2009statistical}, a mathematical approach that models the shape variation of an anatomy of interest in a population \cite{rodero2021linking}. This technique includes statistical shape modelling (SSM), a method to represent the shape probability distribution by a mean shape and modes \textcolor{gr}{describing} the shape variations \cite{biglino2017computational}.  Principal component analysis (PCA) is usually the unsupervised technique used to extract the linearly independent components describing the variation of the shape in \textcolor{gr}{a} population. PCA requires datasets containing the same number of points. \textcolor{gr}{Therefore,} when working with computational grids, iso-topological meshes (i.e. having the same number of nodes and the same connectivity) are required. To this end, radial basis function (RBF) mesh morphing can be used to adapt a reference mesh to a new patient's anatomy \cite{biancolini2017fast,biancolini2020fast}. SSM \textcolor{gr}{has been widely used in literature} for many medical purposes \cite{taghizadeh2019automated,catalano2022atlas,lotjonen2005artificial,ambellan2019statistical}.
A shape-based framework to identify diseased or healthy anatomical structures was described by Durrleman et al. \cite{durrleman2014morphometry}. Bruse et al. performed hierarchical clustering on a set of aortic segmentations and on the reduced PCA-derived shapes to replicate the diagnoses given by clinical experts  \cite{bruse2017detecting}.
\textcolor{gr}{An alternative method to PCA for performing} SSA is partial least squares (PLS) analysis. PLS is a supervised statistical method used to analyze the relationship between two sets of variables, one set of predictor variables and one set of response variables \cite{brereton2014partial}.
It was used and compared with PCA to predict the risk of aortic dissection \cite{williams2022aortic}, showing \textcolor{gr}{superior capabilities} in separating patients who will experience dissection and patients who will not. 
Moreover, it has been used to assess the risk of myocardial infarction and predict cardiac remodelling \cite{suinesiaputra2017statistical,lekadir2016statistical,mansi2011statistical}.

The possibility of combining shape modes and regression models to predict the risk of AsAA surgery was already presented by Cosentino et al. \cite{cosentino2020statistical}. Liang et al. \cite{liang2017machine} used regression methods combined with machine learning models to investigate the relationship between shape features of the thoracic aorta and rupture risk previously derived from numerical simulation. Meyrignac et al. \cite{meyrignac2020combining} combined abdominal aortic lumen volume and parameters derived from numerical simulation such as wall shear stress (WSS) with regression models in order to predict the abdominal aneurysm growth.

\textcolor{gr}{In this paper, using the geometric decomposition techniques presented in our previous work \cite{geronzi14assessment}, we first derive a set of local shape features using a more extensive dataset than the one previously exploited.} After, we perform statistical shape analysis to extract global shape features from principal component analysis and partial least squares.  \textcolor{gr}{Local shape features are geometric measurements of specific segments or curves related to the aneurysm domain whereas global shape features describe the overall shape and dimension of the aneurysm and are extracted using all information related to the studied population.} Finally, we apply regression methods with the aim of directly inferring the ascending aortic aneurysm growth rate based on the information derived from the previously computed shape features.

\section{Materials and Methods}
\begin{figure}[h!]
\begin{center}
\includegraphics[width=\textwidth]{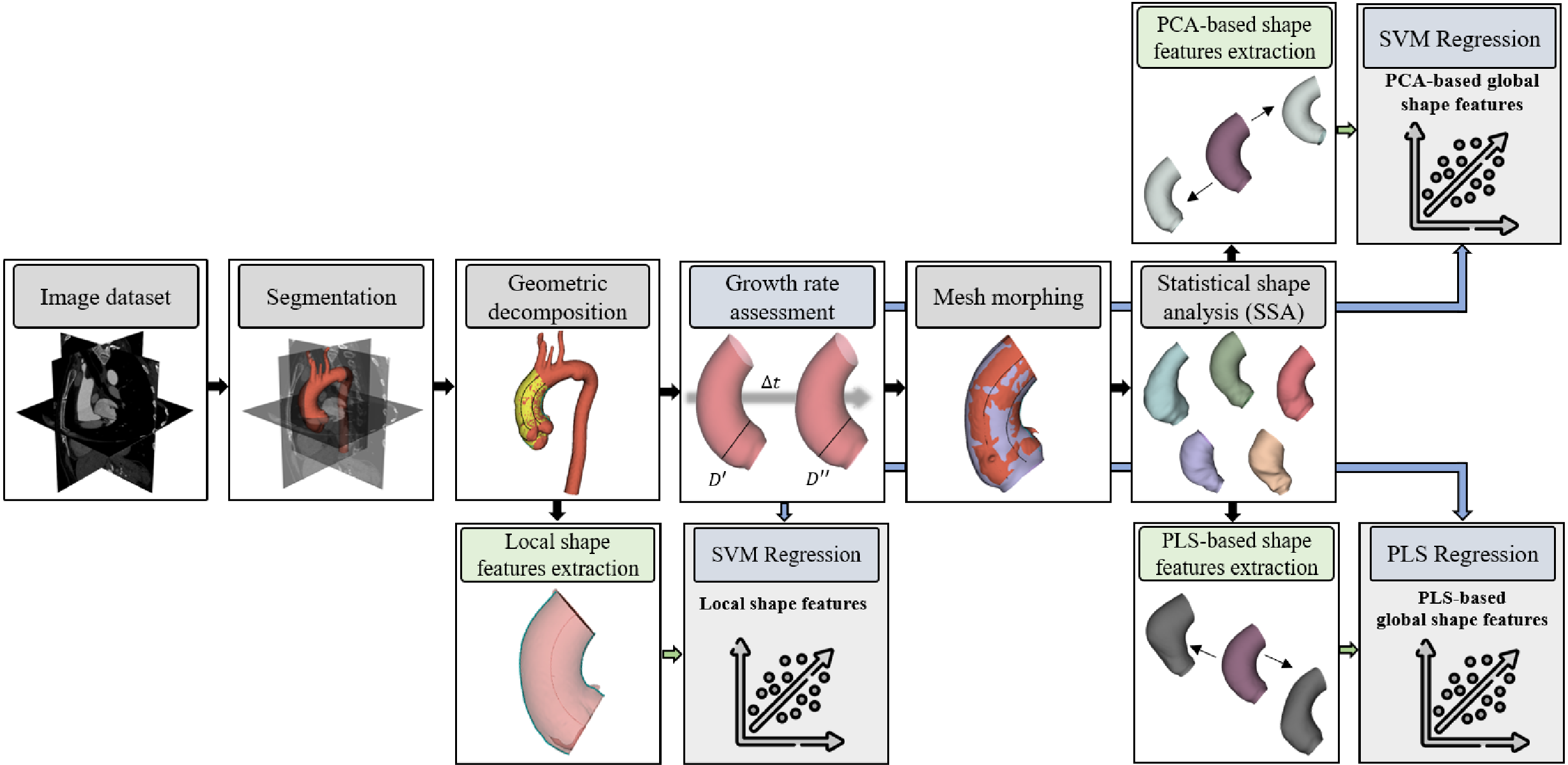}
\end{center}
\caption{Pipeline of the full procedure to assess the growth rate from local and global shape features.} \label{fig:1}
\end{figure}

The principal purpose of this paper is to compare the prediction \textcolor{gr}{performances} of regression techniques applied to assess the ascending aortic aneurysm growth rate through local or global shape features.

The workflow, shown in Fig. \ref{fig:1}, consists of the following steps: (1) selection of a longitudinal dataset consisting in patients with ascending aortic aneurysm; (2) image segmentation to extract 3D geometries; (3) geometric decomposition of the aortic domain, (4) determination of local shape features; (5) computation of the aneurysm growth rate; (6) mesh morphing application to obtain iso-topological surface meshes; (7) implementation of statistical shape analysis based on unsupervised principal components analysis to extract global shape features due to the geometrical variability; (8) execution of statistical shape analysis based on supervised partial least squares to extract global shape features related to the growth of the aorta; (9) assessment of the relationship between local-global shape features and growth rate through regression models. These points are explained in detail below.

\subsection{Dataset} \label{sec1}
We included patients affected by ascending aortic aneurysm with two 3D exams separated by at least 6 months. The dataset was composed of subjects with both bicuspid and tricuspid aortic valve and was provided by three French centres: the University Hospital of Rennes, the University Hospital of Dijon and the University Hospital of Toulouse. The ethical standards were respected in performing this study.
Images from both CT and MRI-Angiography were used, with a maximum voxel size of 1mm x 1mm x 1mm. To increase the likelihood of producing reliable and reproducible results, additional exclusion criteria were set: (1) patients under 25 years of age, (2) aneurysms related to systemic inflammatory diseases, (3) prior aortic valve surgery, (4) acute aortic syndrome, (5) congenital tissue disorders such as Marfan syndrome and (6) images damaged by artefacts.
\textcolor {gr}{In total, $N$ = 70 patients were included, 50 of whom had already been used in our previous work related to the identification of local shape features \cite{geronzi14assessment}. 47 (67.1\%) patients had double ECG-gated acquisitions.} The longitudinal dataset consisted of 120 (85.7\%) CT-Scans and 20 (14.3\%) MRI-Angiographies. 85 acquisitions (60.7\%) were performed with contrast agent injection and 35 (39.3\%) without. 

\subsection {Segmentation} \label{sec2}
\begin{figure}[h!]
\begin{center}
\includegraphics[width=100mm]{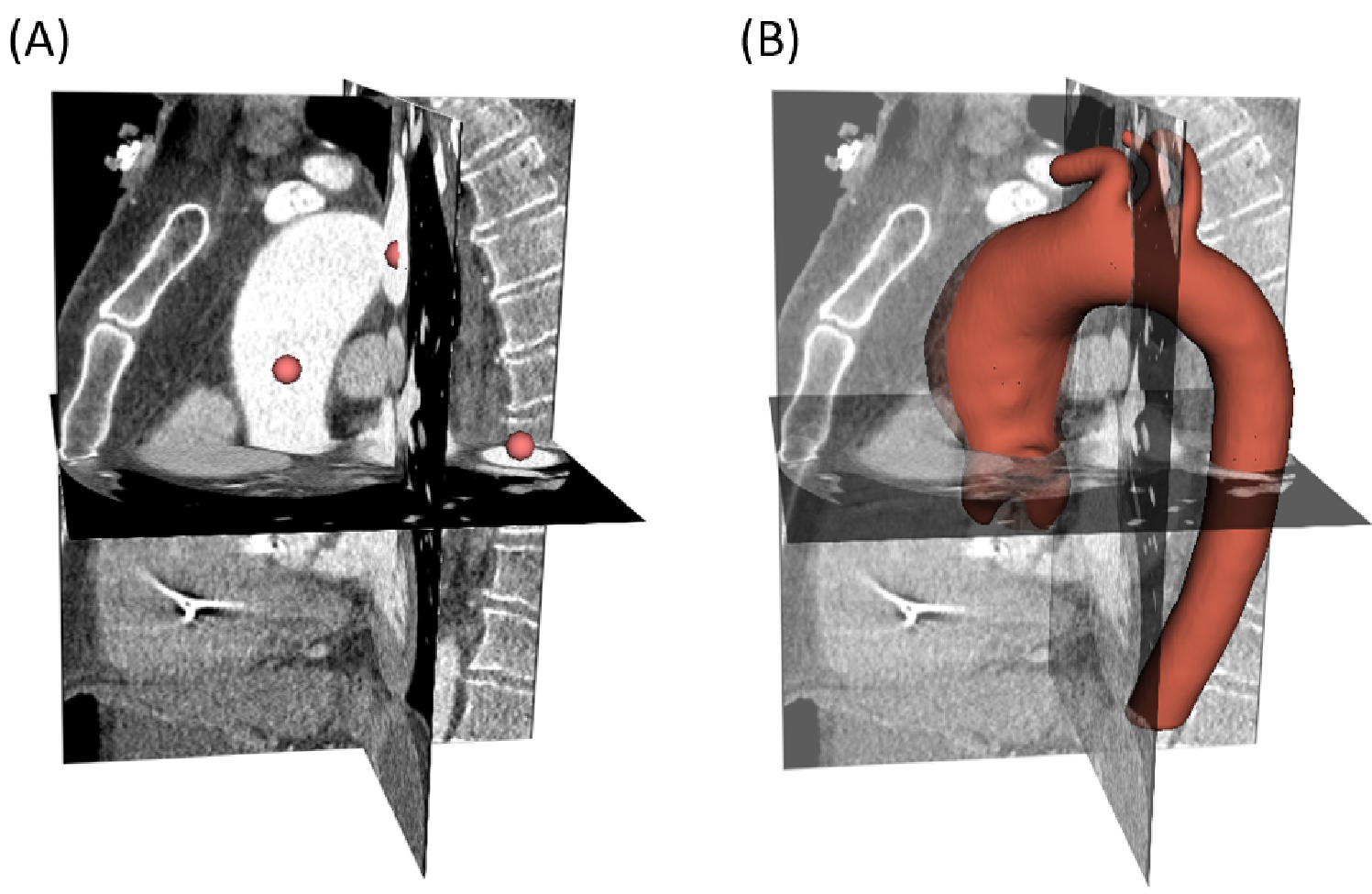}
\end{center}
\caption{(A) Spheres built around the markers to acquire the grey levels for the thresholding method. (B) Final model derived from the segmentation.} \label{fig:2}
\end{figure}

The aortic lumen geometries were extracted from CT and MRI images and the reconstructions were made in 3D Slicer \cite{kikinis20133d}. \textcolor{gr}{We used a semi-automatic local thresholding method based on the grey-level histogram derived from analysing three sets of voxels. Each set was determined by initially identifying three points in different areas of the aorta: the ascending aorta, the aortic arch and the descending aorta. Around each point used as center, three spheres of radius 5 mm were built, as shown in Fig. \ref{fig:2} (A). All the voxels  distributed inside each sphere were acquired to determine the grey level interval for segmenting the aorta.} \textcolor{gr}{Automatic morphological erosion and dilation were applied. Through erosion, 2  voxels were removed from the boundaries of the binary labelmap representing the aorta. A subsequent dilation added the same number of voxels to preserve the original size. Shrinking the domain, erosion was useful for removing small white noises and detaching small connected objects while dilation was applied to restore the overall shape of the vessel.} The resulting surface underwent a manual editing process for the correction of possible improperly segmented portions. A median filter was then applied with a kernel size of 3 mm. Any remaining artefact in the resulting aortic surface (Fig \ref{fig:2} (B)) was removed using Meshlab \cite{cignoni2008meshlab}.

\subsection{Geometric decomposition and local shape features} \label{sec3}
\begin{figure}[h!]
\begin{center}
\includegraphics[width=\textwidth]{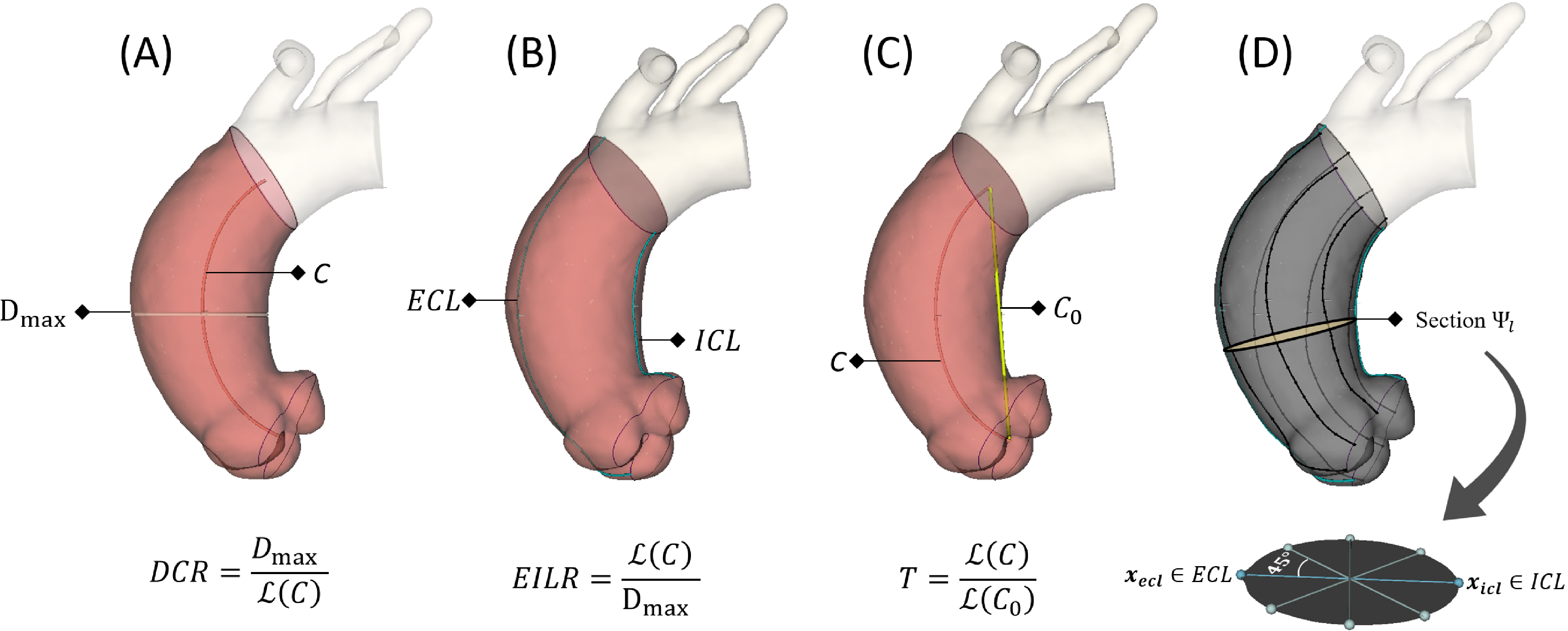}
\end{center}
\caption{(A-C) Local shape features for the ascending aorta model; (D) above, the model with the lateral splines on the surface. Below, a section $\Psi_l$ of the aorta for computing the points that identified the splines.} \label{fig:3}
\end{figure}

\textcolor{gr}{The vessel centerline was extracted using the maximally inscribed sphere method and Voronoi diagrams \cite{antiga2003centerline}.}
The ascending tract was isolated cutting the vessel perpendicularly to the centerline at two specific points: at the level of the annulus and at the end of the ascending aorta where the first ostium of the brachiocephalic artery was detected. The superior branches, the coronary arteries, the arch and the descending aorta were excluded. A spline corresponding to the internal curvature line ($ICL$) was determined from the internal geodesic. The external curvature line ($ECL$) was computed by projecting each point generating the $ICL$ onto the external part of the aorta in a direction perpendicular to the centerline and passing through it \cite{geronzi14assessment}.
Using the geometries derived from the first acquisition, the following local shape features, shown in Fig. \ref{fig:3} (A-C), were extracted:
\begin{enumerate}
\item the diameter-centerline ratio $DCR$:
\begin{equation}
DCR=\frac{D}{\mathcal{L}(C)}
\end{equation}

where $D$ was the maximum diameter and $\mathcal{L}(C)$ was the length of the centerline $C$ related to the ascending tract;

\item the ratio between the external and internal curvature line lengths:
\begin{equation}
EILR=\frac{\mathcal{L}(ECL)}{\mathcal{L}(ICL)}
\end{equation}

\item the tortuosity $T$: 

\begin{equation}
T=\frac{\mathcal{L}(C)}{\mathcal{L}(C_{0})}
\end{equation}

where $C_{0}$ was the straight line connecting the first and the last points of the centerline $C$.
\end{enumerate}

In parallel to the extraction of the local features, some additional splines on the aortic surface were extracted, as shown in Fig. \ref{fig:3} (D). These were subsequently used to obtain the points acting as pseudo-landmarks \cite{frangi2002automatic} to perform mesh morphing. On each aortic section $\Psi_{l}$  ($l \in {1,...,100}$) perpendicular to $C$, given the axis along the direction connecting the point of the centerline to $\mathbf{x_{ecl}} \in ECL$, 6 additional points were identified as the intersection between the axis rotated onto $\Psi_{l}$ each time by 45 degrees and the boundary of the section itself. When connecting the corresponding points on the 100 sections, 6 new splines were thus derived.

\subsection{Growth rate assessment} \label{sec6}
Although the diameter threshold for elective surgery may not have been met, patients experiencing rapid AsAA growth should be carefully and frequently monitored for preventive surgical purposes \citep{geisbusch2014prospective}. For these reasons, we can deduce that the risk of aneurysm rupture is closely related to the risk of aneurysm growth \citep{davies2002yearly}.
By using longitudinal data and naming $D_{i}'$ the diameter $D$ related to the first acquisition of the $i$-patient and $D_{i}''$ the maximum diameter of the model from the follow-up exam, we derived the aneurysm growth rate GR$_i$ by dividing the difference in maximum diameters by the time gap $\Delta \tau_{i}$, measured in months, between the two acquisitions:
\begin{equation}
\textrm{GR}_{i}= \frac{D_{i}''-D_{i}'}{\Delta \tau_{i}}
\end{equation}

All the growth rates were computed and stored in the vector \textbf{Y}:
\begin {equation}
\textbf{Y}=\left(\textrm{GR}_{1}, \textrm{GR}_{2}, \ldots, \textrm{GR}_{N}\right) \in \mathbb{R}^{N}
\end {equation} 
The study about the correlation between the local shape features here proposed and the aneurysm growth rate has already been performed in a previous work \cite{geronzi14assessment}.

\subsection{Mesh morphing }\label{sec4}
The iso-topological grids required for the statistical shape analysis were built using RBF mesh morphing, whose mathematical background is given in Appendix A. The cubic kernel $\varphi(r)=r^3$  was chosen to interpolate the displacements in 3D space \cite{cella2017geometric}. 

The initial shape used to generate the first mesh was identified as the one reporting the median aortic diameter of our patient population, as done for the femur by Grassi et al. \cite{grassi2011evaluation}. The mesh, consisting of  $E$ = 37400 quadrilateral elements and $K$ = 37620 nodes, was obtained using ANSA pre-processor (BETA CAE Systems, Switzerland). A preliminary step was performed to align all the segmented models to the baseline mesh through an iterative closest-point algorithm. As already done by Marin et al. \cite{marin2023segmentation}, a two-step morphing procedure was applied to modify each time the reference mesh in order to exactly match the target segmentation. The first step of the morphing procedure consisted in approaching the target segmented surface and the second in completely projecting the deformed surface on the target geometry to achieve a perfect fit.
Controlling the mesh using morphing is particularly difficult in case of biological models with few detectable anatomical landmarks \cite{vos2004statistical}, as the ascending aorta. In this regard, we developed a method to  extract some pseudo-landmarks \textcolor{gr}{from the 3D surface}, avoiding the need for manual landmark placement. The Source Points (SPs) to drive the morphing, corresponding to the pseudo-landmarks, were automatically derived through an equally-spaced sampling of the previously obtained splines. 10 SPs per spline were collected on the initial model for a total of 80 SPs. \textcolor{gr}{A displacement was imposed to the SPs of the initial model in order to match the SPs extracted from the target geometry} and the mesh nodes were updated through RBF interpolation, as reported in the appendix. To ensure the overlapping of the entire wall, the modified surface nodes were projected onto the target segmentation in the second step. The direction of projection was determined by the normal of each node of the reference mesh.
To reduce mesh distortion due to morphing, once the $N$ = 70 iso-topological grids were obtained, a mean template was derived and mesh morphing was performed again on all grids starting from this new average model. A new mean template was then generated and used for the subsequent steps.

\subsection{Statistical shape analysis} 
Exploiting the set of iso-topological grids, a data matrix \textbf{X} containing $K$ shape vertices was created:

\begin {equation}
\textbf{X}=\left(\mathbf{x_1}, \mathbf{x_2}, \ldots, \mathbf{x_N}\right) \in \mathbb{R}^{3 K \times N}
\end {equation} 

The statistical shape analysis to extract global shape features was performed both by creating a statistical shape model based on the principal components and using partial least squares analysis. All the algorithms were developed using Python.

\subsubsection{Statistical shape model}

Principal component analysis was used to extract the principal modes of variation by computing the eigenvectors of the covariance matrix \textbf{C} of the training data:
\begin {equation}
\label{cov}
\textbf{C}=\frac{1}{N-1} \textbf{X}\textbf{X}^T \in \mathbb{R}^{3 K \times 3 K}
\end {equation}
The eigen-equation related to the covariance matrix is:
\begin{equation}
\textbf{C} \boldsymbol{\phi_j}=\lambda_j \boldsymbol{\phi_j}
\end{equation}

where $\boldsymbol{\phi_j}$ is the eigenvector corresponding to the eigenvalue $\lambda_j$ and represents the directions of variation of the data. Eigenvalues and eigenvectors are ordered from high to low variance. The contribution of each shape mode to the total variance was given by its corresponding eigenvalue $\lambda_j$ \cite{davies2002minimum}.

A factorization of the data matrix through singular value decomposition (SVD) can be performed:

\begin{equation}
\textbf{X} =\textbf{USV}^T
\end{equation}
with $\textbf{U} \in \mathbb{R}^{3 K x 3 K}$ and $\textbf{V} \in \mathbb{R}^{N x N}$  unitary matrices ($\textbf{U}^{-1}=\textbf{U}^* \text { and } \textbf{V}^{-1}=\textbf{V}^*$) and $\textbf{S} \in \mathbb{R}^{3 K x N}$ matrix containing the singular values $s_j$ on the diagonal.

 Thus, equation \ref{cov} can be written as:
 \begin{equation}
\textbf{C} =\frac{1}{N-1}\left(\textbf{USV}^T\right)\left(\textbf{USV}^T\right)^T
\end{equation}
and simplifying:
 \begin{equation}
\textbf{C} =\frac{1}{N-1} \textbf{US}^2\textbf{U}^T
\end{equation}

This demonstrates that the singular values of the data matrix are related to the eigenvalues of the covariance matrix:
\begin{equation}
\lambda_j=\frac{1}{N-1} s_j^2
\end{equation}

Once the template or mean shape $\overline{\boldsymbol{x}}$ has been extracted, each patient shape $\boldsymbol{\tilde{x_i}}$ belonging to the dataset can be reconstructed using the first $M$ shape modes:

\begin{equation}
\boldsymbol{\tilde{x_i}}(\boldsymbol{w})=\bar{\boldsymbol{x}}+\boldsymbol{\phi} \boldsymbol{w}
\end{equation}
where $\boldsymbol{w}$ is the vector containing the shape feature weights for the i-patient which can be derived from:

\begin{equation}
\boldsymbol{w}=\boldsymbol{\phi}^T(\boldsymbol{x_i}-\bar{\boldsymbol{x}})
\end{equation}

Assuming the data follows a normal distribution, each feature weight $w_{j}$ is conventionally bounded within a certain range of the standard deviation:

\begin{equation}
-\xi_{lim} \sqrt{\lambda_j} \leq w_{j} \leq \xi_{lim} \sqrt{\lambda_j}
\end{equation}
where $\xi_{lim}$ is usually assumed equal to 3. In other words, $w_{{j}}$ is a scalar value providing the geometrical influence of each shape mode on the final deformed model.
Only the first $M$ of the $N$ eigenvectors were selected to account for a predetermined percentage of the variance and synthetically represent each aortic shape in the dataset. 
$M$ can be chosen by computing the compactness (CN) and finding the number of shape modes for which the variance curve reaches 80\%, 90\%, 95\% or 99\%. CN is defined as the sum of variances normalized by the whole cumulative variance:

\begin{equation}
\textrm{CN}(M)=\frac{\sum_{j=1}^{M} \lambda_j}{\sum_{j=1}^{N} \lambda_j}
\label{Compactness}
\end{equation}

The CN curve shows how many PCA modes are needed to describe a certain amount of variation in the dataset. A second parameter to assess the SSM quality is the generalization (GE). It is used to estimate the SSM capability to represent unseen data and is computed as the average sum of square errors of a leave-one-out (LOO) procedure \cite{wong2015performance}. Each time, in fact, one patient is excluded and a new statistical shape model is built using the $N-1$ remaining ascending aortic shapes. The new SSM is then used to reconstruct the shape of the left-out patient and the difference between the original shape and the reconstruction is quantified using the mean square error, progressively including additional modes. In this case, GE was computed using up to $M$ shape modes:

\begin{equation}
\textrm{GE}(M)=\frac{1}{N} \sum_{i=1}^{N}\left\|\mathbf{x_i}-\hat{\mathbf{x_i}}\left(M\right)\right\|^2
\label{Generalization}
\end {equation}
where $\mathbf{x_i}$ and $\hat{\mathbf{x_i}}$ are the original and rebuilt left-out shapes, respectively.

\subsubsection{Partial least squares analysis}
PCA modes are extracted purely from the patient matrix \textbf{X} without taking into account any external information related to the examined shapes.
On the other side, PLS performs a simultaneous decomposition of \textbf{X} and \textbf{Y} in order to obtain the highest correlation for the score vectors of both the input and output matrices \cite{helland1988structure}. This ensures maximal interdependencies between the shapes and the output variables, making the statistical shape decomposition application-oriented since \textcolor{gr}{dependent on} the clinical response variables, i.e. the growth rate. 
Whereas PCA tries to identify hyperplanes that capture the most significant variation in the data, PLS employs a linear regression model that involves projecting the predicted variables and observable variables into a new space to establish the fundamental relationships between them. PCA generates a set of orthogonal components that are uncorrelated and ordered by the amount of variance. PLS, on the other hand, generates a set of latent variables that capture the maximum covariance between the \textbf{X} and \textbf{Y}  matrices.

Given two standardized matrices $\textbf{X} \in \mathbb{R}^{N \times 3K}$ and $Y \in \mathbb{R}^{N \times Z}$ where $N$ is the number of observations (shapes), $3K$ is the number of predictor variables (point coordinates), $Z$ is the number of predicted variables and defining the number of shape modes $M$, PLS models the relations between these two matrices through score vectors. It decomposes the \textbf{X} and the \textbf{Y} matrices as follow:

\begin{equation}
\label{general_eq_PLS}
\begin{aligned}
& \mathbf{X}=\mathbf{T P}^T+\mathbf{E} \\
& \mathbf{Y}=\mathbf{U} \mathbf{Q}^T+\mathbf{F}
\end{aligned}
\end{equation}

where  $\mathbf{T} \in \mathbb{R}^{N \times M}$, $\mathbf{U} \in \mathbb{R}^{N \times M}$ are the matrices of the $M$ extracted score vectors \textbf{t} and \textbf{u}, $\mathbf{P} \in \mathbb{R}^{3K \times M}$ and $\mathbf{Q} \in \mathbb{R}^{Z \times M}$ represent the matrices of loadings and $\mathbf{E} \in \mathbb{R}^{N \times 3K}$ and $\mathbf{F}\in \mathbb{R}^{N \times Z}$ are the matrices of residuals. The PLS method finds weight vectors $\mathbf{b}$, $\mathbf{c}$ such that:
\begin{equation}
[\operatorname{cov}(\mathbf{t}, \mathbf{u})]^2=[\operatorname{cov}(\mathbf{X b}, \mathbf{Y c})]^2=\max _{|\mathbf{r}|=|\mathbf{s}|=1}[\operatorname{cov}(\mathbf{X r}, \mathbf{Y s})]^2
\end{equation}

where $\operatorname{cov}(\mathbf{t}, \mathbf{u})=\mathbf{t}^T \mathbf{u} / N$ denotes the sample covariance between the score vectors.

PLS is based on an iterative process: the nonlinear iterative partial least squares (NIPALS) algorithm \cite{wold1975path}. It starts with a random initialization of the score vector \textbf{u} and executes the following steps until convergence is reached:

\begin{itemize}
\item $\mathbf{b}=\mathbf{X}^T \mathbf{u} /\left(\mathbf{u}^T \mathbf{u}\right)$
\item $\|\mathbf{b}\| \rightarrow 1$
\item $\mathbf{t}=\mathbf{X} \mathbf{b}$
\item $\mathbf{c}=\mathbf{Y}^T \mathbf{t} /\left(\mathbf{t}^T \mathbf{t}\right)$
\item $\|\mathbf{c}\| \rightarrow 1$
\item $\mathbf{u}=\mathbf{Y} \mathbf{c}$
\end{itemize}

Since in this case $Z=1$, $\mathbf{Y}$ can be denoted as $\mathbf{y}$  and $\mathbf{u}=\mathbf{y}$. Consequently, the NIPALS procedure converges in a single iteration.
The weight vector \textbf{b} is equal to the first eigenvector of the following eigenvalue problem \cite{hoskuldsson1988pls}:

\begin{equation}
\mathbf{X}^T \mathbf{Y} \mathbf{Y}^T \mathbf{X} \mathbf{b}=\lambda \mathbf{b}
\end{equation}

After the extraction of the score vectors \textbf{t} and \textbf{u}, a process of deflation of the matrices \textbf{X} and \textbf{Y} is performed by subtracting their rank-one approximations based on $\mathbf{t}$ and $\mathbf{u}$.
Various methods of deflation are used, which define different versions or variants of PLS.
The vectors of loadings $\mathbf{p}$ and $\mathbf{q}$ can be derived from (\ref{general_eq_PLS})  as coefficients of regressing $\mathbf{X}$ on $\mathbf{t}$ and $\mathbf{Y}$ on $\mathbf{u}$, respectively:

\begin{equation}
\mathbf{p}=\mathbf{X}^T \mathbf{t} /\left(\mathbf{t}^T \mathbf{t}\right) \text { and } \mathbf{q}=\mathbf{Y}^T \mathbf{u} /\left(\mathbf{u}^T \mathbf{u}\right)
\end{equation}
Since $Z=1$,  the PLS1 deflection method can be used:

\begin{equation}
\mathbf{X}=\mathbf{X}-\mathbf{t p}^T 
\end{equation}
It is based on the assumption that the score vectors $\mathbf{t}$ are good predictors of $\mathbf{Y}$ and that a linear inner relation between the scores vectors $\mathbf{t}$ and $\mathbf{u}$ exists, i.e:

\begin{equation}
\mathbf{U}=\mathbf{T D}+\mathbf{H}
\end{equation}
where $\mathbf{D} \in \mathbb{R}^{M \times M}$ is the diagonal matrix and $\mathbf{H}$ denotes the matrix of residuals,
The deflation of \textbf{y} is technically not needed in PLS1.

For the PLS modes, the new patient-specific shape features score vectors $\mathbf{t_{_{i}}}$ were computed and used for the prediction.

\subsection{Regression} \label{sec5}
Once the shape features have been computed, regression models were used to directly infer the patient-specific growth rate. Our approach involved both local and global shape features. Regarding the local, $DCR$, $EILR$ and $T$ were employed together. For the global, the patient-specific weights $\mathbf{w}$ related to the PCA modes and the PLS scores $\mathbf{t}$ were used as a predictor of the growth rate.
Concerning the PCA modes, we used a F-test as feature ranking algorithm to order the predictors by importance \cite{mahfouz2007patella}. Higher scores were associated with higher-importance shape features. The null hypothesis of each F-test is that the means of the response values, which are grouped by predictor variable values, are drawn from populations with equal means. On the other side, the alternative hypothesis is that the means of the populations are not all the same. If the resulting p-value of the test statistic is small, the corresponding predictor variable has a significant impact on the response variable. We reported as output the scores of the F-test $FS=-log(p)$. Thus, a high score value indicates that the corresponding predictor is relevant. The first three ordered PCA modes were used as global shape features. Although not necessary for the PLS shape features choice, the same F-test was applied to the PLS scores to observe the differences with the scores from PCA.
The regression model used for the local shape features and the PCA-based global shape features was the support vector machine (SVM), which had already shown good results in similar studies \cite{liang2017machine}. This machine learning model is able to describe the nonlinear relationships between shape features and aneurysm growth. A gaussian kernel function was set and the hyperparameters, reported in the results section, were tuned to minimize the prediction mean square error (MSE) \cite{jeng2006hybrid}. Regarding the PLS regression, we only considered the first three components of the PLS, as these refer to the shape features most significantly correlated with the computed growth rate. 
Leave-one-out cross-validation was performed to evaluate the performance of each regressor and the regression accuracy was determined through root mean square error (RMSE). R$^2$ values are reported for both local and global shape features to assess the regression fit quality. The marginal effect of each predictor on the response in terms of growth rate is described by reporting the partial dependence plots between the predictor variables and the predicted responses. 
Finally, the SVM regression surface for the representative PCA-based global shape features is provided to understand how the predicted growth rate varied as the shape changed according to the variability of the studied population.

\section{Results}
The population mean age at the baseline acquisition was 62.7 $\pm$ 15.5 years. 21 women (30\%) and 49 men (70\%) were included in the dataset. The mean time between two consecutive acquisitions was 18 $\pm$ 16 months, with a range between 6 and 98 months. At the baseline, the maximum diameter was 49.4 $\pm$ 4.1 mm whereas at the follow-up it was 51.9 $\pm$ 4.6 mm. The full dataset presented a median growth rate of 0.118 mm/month with an interquartile range IQR = 0.133 mm/month. 

The local shape features reported in terms of mean value and standard deviation were: $EILR = 2.336\pm0.323$, $CDR = 0.496\pm0.052$ and $T = 1.213\pm0.072$. 

Mesh morphing was applied on all geometries, ensuring a correct correspondence between the pseudo-landmarks computed on the template model and the pseudo-landmarks identified on the target geometry after the first step and a perfect matching after the second.

\begin{figure}[h!]
\begin{center}
\includegraphics[width=\textwidth]{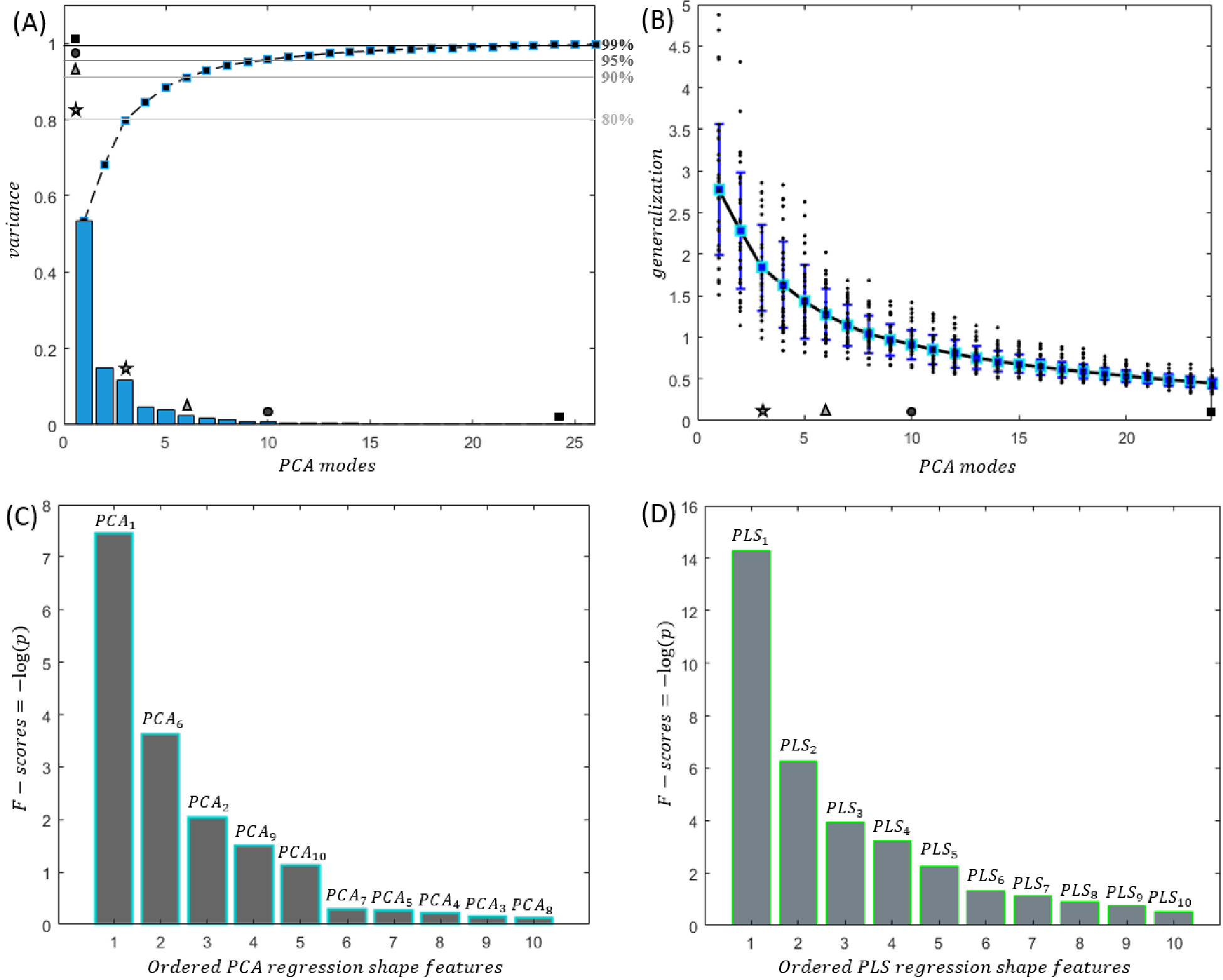}
\end{center}
\caption{(A) SSM compactness according to the number of PCA modes used: the symbols indicate the PCA mode for which 80\%, 90\%, 95\% and 99\% of the cumulative variance is reached. (B) Generalisation curve when increasing the number of PCA modes. (C) $FS$ for selecting the PCA modes to perform SVM regression. (D) $FS$ for the PLS modes.}
\label{fig:4}
\end{figure}

The statistical shape analysis was performed including all computational surface grids.

\begin{figure}[h!]
\begin{center}
\includegraphics[width=11.3cm]{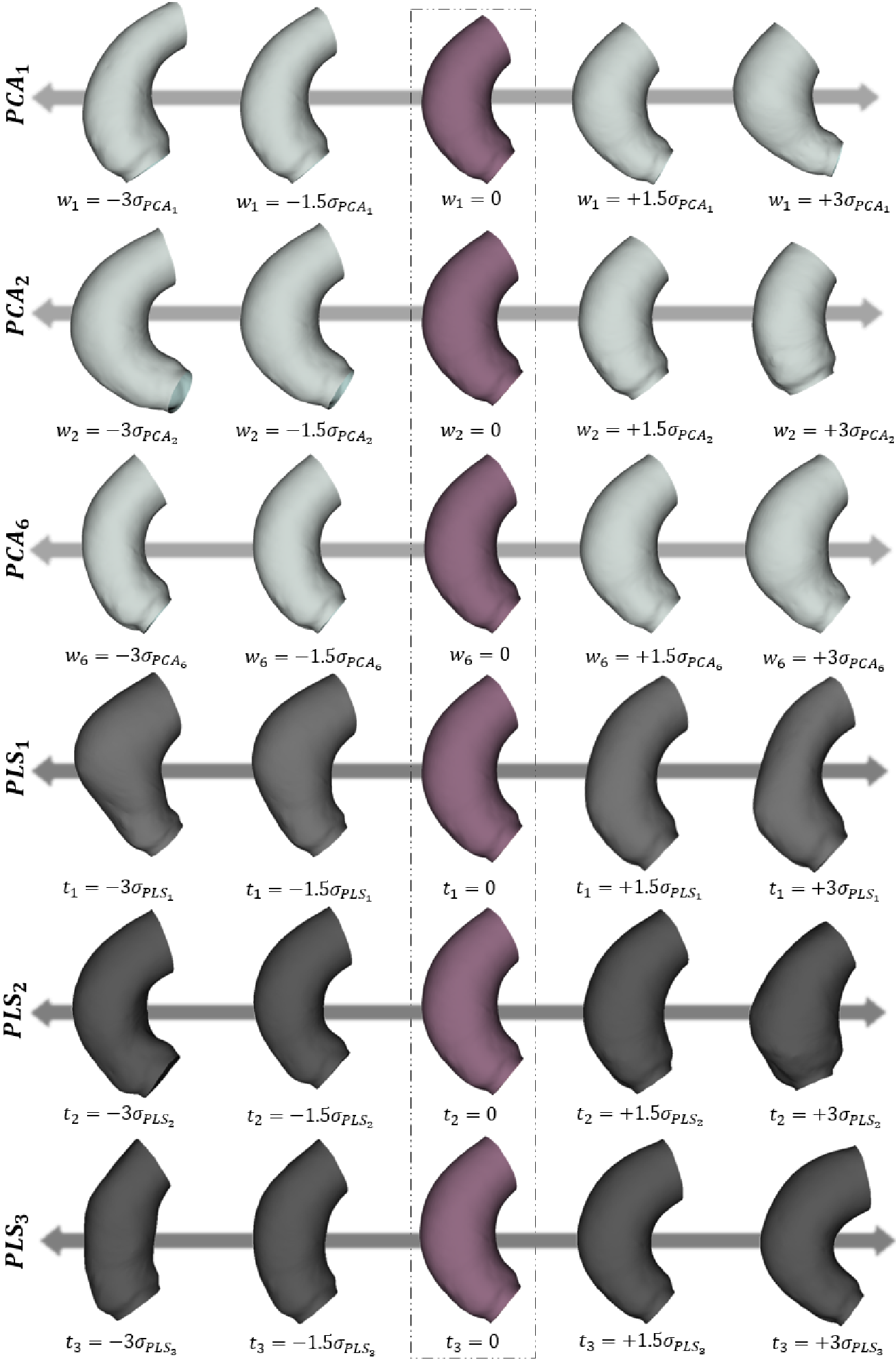}
\end{center}
\caption{Shape modification due to PCA modes 1, 2 and 6 and PLS modes 1, 2 and 3.} \label{fig:5}
\end{figure}

Concerning the SSM, the compactness curve is reported in Fig. \ref{fig:4} (A). The first PCA mode alone accounted for 52.4\% of the anatomical variability in the population whereas the first 3 PCA modes together captured 80\% of the overall variability. 90\%, 95\% and 99\% of the compactness curve were achieved using 6, 10 and 24 PCA modes, respectively. The generalization ability is instead reported in Fig.
\ref{fig:3} (B) for the first 24 PCA modes. By including additional shape modes, its mean value, representing the mean square error between shapes reconstructed by LOO and shapes reconstructed by the SSM over the entire population, went from 2.81 $\textrm{mm}^2$ to 0.52 $\textrm{mm}^2$, where it tends to stabilise. 
In Fig. \ref{fig:4} (C), the results of the F-test for choosing the PCA modes to be used for the regression are reported. Based on $FS$, the PCA modes selected as global shape features were 1, 2 and 6. Fig. \ref{fig:4} (D) shows the same outputs for PLS score vectors $\mathbf{t}$. 
 Concerning the first 10 shape modes, PLS globally reported higher $FS$ values than PCA. 
The contribution of each mode can be visualized by deforming the mean template from low ($\xi=-\xi_{lim}$) to high ($\xi=+\xi_{lim}$) standard deviation, as reported in Fig. \ref{fig:5} for the three selected PCA and PLS modes. 

The first PCA mode defines the overall position of the aneurysm in the ascending tract. Negative weight values are associated with aneurysms  located closer to the root whereas positive weight values define aneurysms developed more towards the end of the ascending aorta. The second PCA mode mainly describes the curvature and tortuosity of the ascending aorta. PCA mode number 6 is instead graphically associated with the size of the aneurysm, which increases as the weight of the mode itself rises. On the other hand, the first PLS mode appears to represent both the location of the aneurysm and its size. The second PLS mode is visually mainly associated with the diameter of the aneurysm while the third mode is graphically related to the tortuosity of the ascending tract.

\begin{figure}[h!]
\begin{center}
\includegraphics[width=\textwidth]{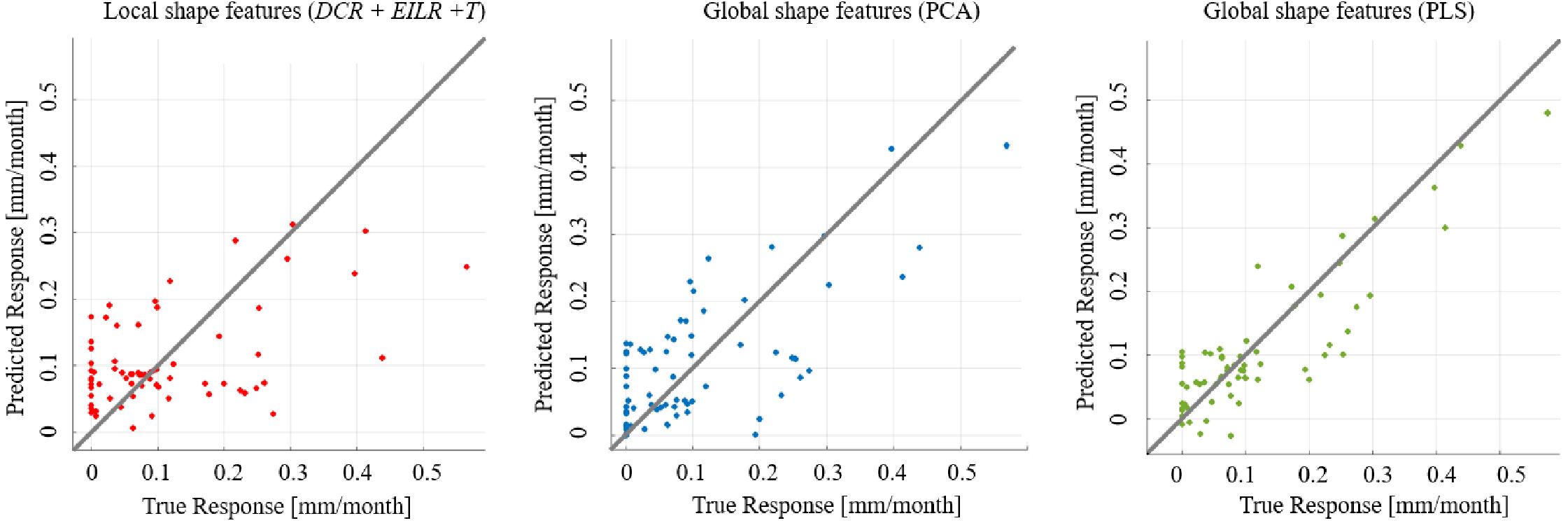}
\end{center}
\caption{Predicted versus true response plot for the growth rate using the three local shape features (red), the global shape features extracted from PCA (blue) and the global shape features extracted from PLS (green).}
\label{fig:6}
\end{figure}

After the LOO cross-validation, the hyperparameters for the SVM regression models are reported in Table \ref{Tab:1}.
The regression performances in terms of R$^2$ and RMSE$_{reg}$ values are the following: R$_{lsf}^2$ = 0.28 and RMSE$_{lsf}$ = 0.112 mm/month,  R$_{PCA}^2=0.42$ and RMSE$_{PCA}$ = 0.083 mm/month and  R$_{PLS}^2$ = 0.63 and RMSE$_{PLS}$ = 0.066 mm/month. The comparison between real and predicted growth rate values for the three regression models is shown in Fig. \ref{fig:6}.

\begin{table}[h!]
\centering
\begin{tabular} {||c| c| c||}
\hline
{Hyperparameters} &  {local shape features} &  {global shape features (PCA)} \\ \hline
{Kernel size}  & 1.72  & 1.89  \\  
{Box constraint}  &  0.41 & 1.82   \\
{Epsilon}  & 0.008  & 0.039  \\
\hline
\end{tabular}
\caption{SVM regression hyperparameters}
\label{Tab:1}
\end{table}

\begin{figure}[h!]
\begin{center}
\includegraphics[width=\textwidth]{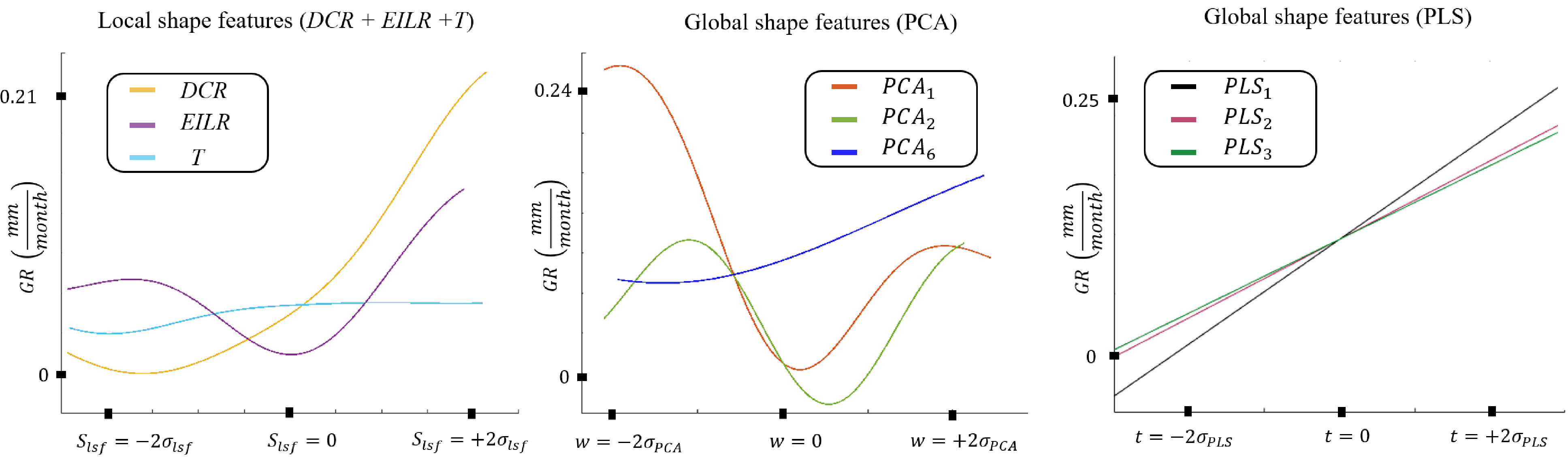}
\end{center}
\caption{Partial dependencies plots for local and global shape features.} \label{fig:7}
\end{figure}

The partial dependencies plots are shown in Fig. \ref{fig:7} for the three cases: the first graph shows the dependence of the gaussian SVM regression model on the three local shape features, the second reports how the selected PCA modes affect the SVM prediction while the third shows the estimated linear relationship between the three PLS modes selected and the growth rate.

Finally, in Fig. \ref{fig:8}, the regression surface is shown using the modes extracted from the statistical shape model. The ascending aorta shapes corresponding to specific zones of the surface are reported to associate the shape with the estimated growth rate.

\section {Discussion}
In this work, we presented a method to exploit local and global shape features for the prediction of the ascending aortic aneurysm growth rate. By using the framework we proposed, based on geometric decomposition, mesh morphing, statistical shape analysis and regression, we were able to extract distinctive shape features potentially valuable for improving the prediction of the aneurysm growth.
The results of this study show that the partial least squares regression model based on global shape features can outperform the support vector machine regression models based on local shape features and global shape features derived from principal component analysis. 

The higher frequency of male patients agrees with what is observed for the aneurysm disease \cite{bossone2021epidemiology}. The obtained growth rate results are consistent with what has been reported in literature \cite{elefteriades2015indications,oladokun2016systematic}. 

Concerning the initial grid, we chose the patient's model reporting the median diameter to reduce the mesh degradation after mesh morphing \cite{geronzi2021high}. This starting model is a reasonable compromise to reach the aneurysms with the smallest and largest diameter in the dataset. \textcolor{gr}{Like the maximally inscribed sphere method, the geometric decomposition allowed to derive the maximum diameter along the vessel centerline but it additionally ensured the possibility of identifying sections perpendicular to the centerline which enable the creation of the splines.} The number of pseudo-landmarks acting as Source Points to perform the first morphing step was chosen in order to achieve good results in terms of computational grid quality \cite{marin2023segmentation}. A fundamental requirement for building accurate statistical shape models is the one-to-one correspondence between the positions of the landmarks on each geometry of the dataset. Since it is very complex to identify landmarks for the isolated ascending tract, the proposed morphing method allows associating points of the splines resulting from the geometric decomposition of the initial template to the same ones computed on the target aorta. This ensured a better control of the grid distortion than using purely distance-based methods in which iterative energies stabilization methods and recursive smoothing techniques, i.e. not driven by statistical information, are usually performed \cite{heimann2009statistical}.

\begin{figure}[h!]
\begin{center}
\includegraphics[width=\textwidth]{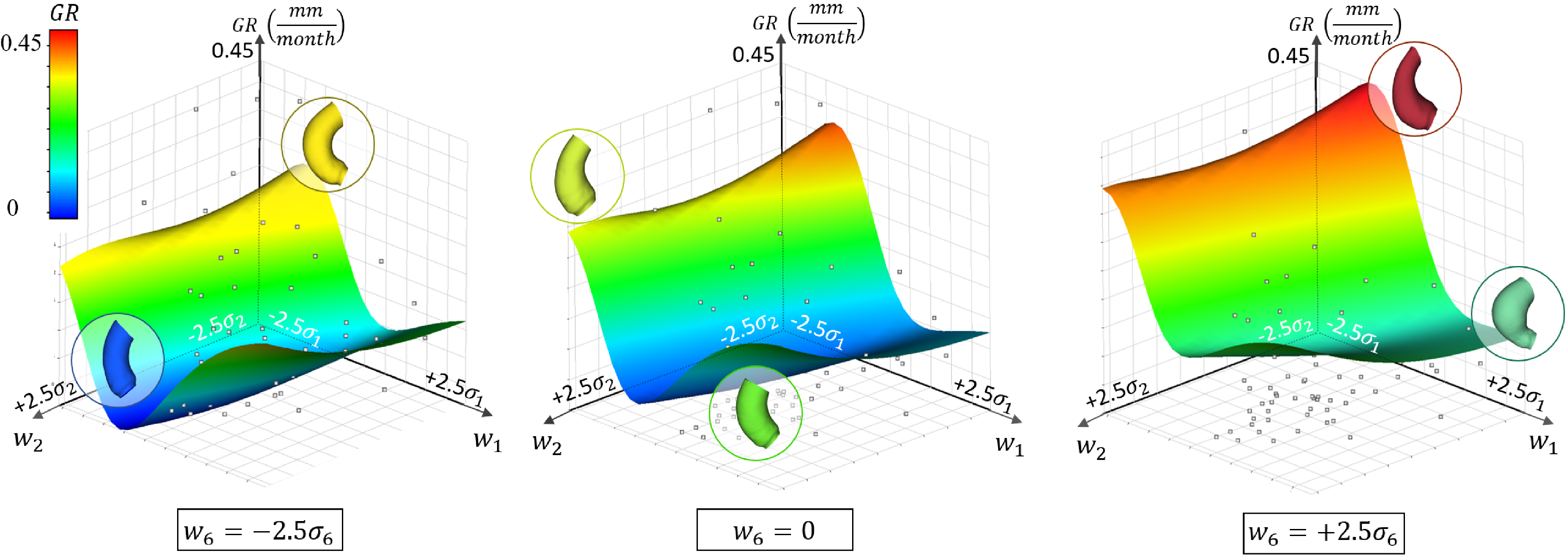}
\end{center}
\caption{Regression surface derived from the global shape features obtained with PCA.} \label{fig:8}
\end{figure}

The SSA was carried out on the ascending tract and not on the entire aorta because a detailed and restricted correspondence between ascending aortic aneurysm shape and growth rate was sought. In this way, any spurious component related to other parts of the thoracic aorta was not included. \textcolor{gr}{While in other approaches \cite{do2018prediction}, PCA modes were related to the cumulative energy with which the aneurysm grew over time for a specific patient, in this work the modes indicated which shape feature within the population could be related to the growth of the disease.} The parametric 3D model offers an advantage compared to 2D metrics extraction in capturing complex ascending aortic shapes by allowing for the detection of shape features that can be represented visually and numerically. These features are difficult to be obtained using conventional morphometric measurements \cite{reutersberg2016measurements}.
Generally, in constructing statistical shape models based on PCA, the high-frequency modes are discarded since considered principally related to noise. However, they could be significant in explaining the pathological growth associated with the disease. Despite including shape modes up to 99\% of the variability, one limitation of PCA-based growth prediction is that there could be an excluded high-frequency mode that is nevertheless strongly associated with growth.
The compactness values obtained for the statistical shape model are in agreement with those indicated by Casciaro et al. \cite{casciaro2014identifying}, in which a healthy subset of aortas required only 6 modes to capture 84\% of the variance, whereas a congenital set of aortas, required 19 modes to capture 90\%. These values are quite consistent with our findings considering that we only selected the ascending part in building the SSM and the variability, in our case, is consequently lower.
Moreover, our compactness and generalization outcomes fit within the range of variability reported in other similar studies \cite{bruse2017detecting,liang2017machine}. This demonstrated the valuable ability of the SSM to represent a wide population.
The F-test results using the PCA components give high importance to shape modes determining the aneurysm location, its size and the tortuosity of the ascending tract. This agrees with the correlation study between local shape features and growth reported in our previous work \cite{geronzi14assessment} and with the shape representation provided by the PLS modes.

Only three global shape features were selected because it was the number for which the root mean square prediction error was the lowest. Using a different number of shape features for the regression, RMSE$_{PCA}$ went from 0.089 mm/month to 0.142 mm/month and RMSE$_{PLS}$ went from 0.066 mm/month to 0.121 mm/month.

Based on this dataset, the value of R$^2_{lsf}$ and RMSE$_{lsf}$ and the representation of real versus predicted response values for local shape features in Fig. \ref{fig:6} indicate that the SVM regression method is  highly inaccurate in predicting the growth rate, in particular for patients whose growth is very rapid, especially if the prediction error is compared to the median GR of the dataset. Results improve when using global shape features.
The RMSE was in fact lowered using SVM with a combination of PCA-based shape features and was further reduced by approximately 56\% from RMSE$_{lsf}$ using PLS regression.  Better results when using global shape features compared to local shape features were already reported by Liang et al. \cite{liang2017machine} for the classification of patients whose aneurysms might burst according to numerical simulation results.
\textcolor{gr}{In addition, it is worth observing that partial dependencies plots (Fig. \ref{fig:7}) for local shape features highlight a major dependence on $DCR$ than $EILR$ and even more than $T$, an aspect already emerged in our previous work \cite{geronzi14assessment}, in which the linear correlation between local shape features and growth rate using the reduced dataset was stronger for the first index.} 

\textcolor{gr}{Our method does not currently allow to derive the shape of the aorta with the associated uncertainty after a certain time interval \cite{do2018prediction} but allows to identify those aneurysm shapes within the population that may evolve most rapidly.} Regarding the three PCA-based global shape features, it is clear that patients with $\xi$ values close to zero are those for whom growth is generally slower. On the other hand, the partial dependencies resulting from PLS show that the tendency to grow is for aneurysms located more towards the root, with a larger initial diameter and for ascending tracts with high tortuosity. The increased risk related to root aneurysms rather than mid-aortic dilatations has already been highlighted by Kalogerakos et al. \cite{kalogerakos2021root}.
In Fig. \ref{fig:8}, the surface resulting from the regression for the three PCA modes is reported. It can be related to what Fig. \ref{fig:7} shows. Fast AsAA growth seems to be related to highly negative $w_{{1}}$ and $w_{{2}}$ and positive $w_{{6}}$.  These results agree with what was found by Della Corte et al. \cite{della2013pattern}: a root phenotype characterized by aortic dilatation at the sinuses may indicate a more severe level of aortopathy. Slow growth occurs instead for aneurysms with values close to 0 for $w_{{1}}$ and $w_{{2}}$ and negative $w_{{6}}$ i.e. for ascending aortas less tortuous, with aneurysms located far from the root and with a smaller initial diameter, results consistent with previous studies \cite{geisbusch2014prospective,van2022ascending}.

Obviously, additional research is needed to understand the connection among aortic shape, wall properties, haemodynamics, mechanical behaviour and aneurysm growth or rupture \cite{garcia2012mechanical}. A more accurate prediction could be probably achieved by combining both shape and physical parameters derived from images or numerical simulation \cite{williams2022aortic}.

This work, therefore, showed the importance of morphometric analysis to improve the prediction of aneurysm growth of the ascending aorta. Combining SSA and regression methods could be a powerful way to model the relationship between shapes and growth rate. Accurately identifying which patients with AsAA will require surgery within a specific timeframe would enhance the risk-benefit analyses and the definition of surveillance protocols \cite{mclarty2015surveillance}. A slow-growing ascending aortic aneurysm, in fact, would not require frequent monitoring whereas it will be necessary for rapid-growing cases.

However, this retrospective study presents some limitations that need to be reported. The principal is the small cohort of patients. The statistical shape analysis requires a large population of representative training samples: for the same topology, the wider the diversity between anatomical  models, the higher the number of samples required. The set of shapes that can be described by the feature space is limited to the deformation modes derived from the included cohort. Therefore, there is no guarantee that a feature vector can accurately represent every new given anatomy. 
\textcolor{gr}{A second significant limitation concerns the linear growth rate hypothesis; exponential growth rate models \cite{martufi2013multidimensional} seem to be more accurate but often require at least three exams over time to be validated \cite{zhang2019patient,prestigiacomo2009predicting} while the dataset selected for this work includes patients with only two acquisitions.}

Uncertainty is then introduced by using both MRI images and CT-Scans. However, we tried to mitigate it by including the resolution criterion and segmenting the intra-luminal aortic region to derive the shape parameters \cite{frazao2020multimodality}.
Moreover, we included patients with non-gated acquisitions, which reduced the accuracy of the results \cite{lehmkuhl2013inter}. This is why an exclusion criterion of 6 months  was set so that \textcolor{gr}{variations were more likely to be attributed} to the growth of the vessel rather than the variations that could emerge between systole and diastole.
Furthermore, we did not consider if the patients, during the follow-up period, had taken drugs such as beta-blockers to slow down the aneurysm growth \cite{goldfinger2014thoracic}, a phenomenon that could alter the calculation of the growth rate.
In performing the regression, a robust study should include a testing dataset but, given the small number of patients available, we preferred to use only the training and validation set. Furthermore, for local shape features and PCA-based global shape features, we have only reported SVM-based regression results. Other regression models should be tested in order to compare the prediction performances.
In our work, we only considered the ascending aorta shape properties and we did not include the patient's valve type although its type and condition can influence haemodynamics and consequently the aneurysm growth rate \cite{girdauskas2016functional}. Many other factors could then be included: aortic annulus disjunction, dislocation of the coronary ostia or possible aortic wall thinning \cite{pisano2020risk}. Moreover, this study does not take into account the arch and the descending tract: anatomical-functional variations of these parts could alter the growth of the ascending aortic aneurysm \cite{suh2014aortic}.
\textcolor{gr}{In future works, blood pressure, flow and tissue material properties and 3 layer-thickness walls will be incorporated into the models to improve the growth prediction. Moreover, numerical simulation will be used to extract biomechanical and hemodynamic   biomarkers which will be complemented with the information provided by the related aneurysm shape features. Parameters such as wall shear stress have, in fact, proven to be good candidates for predicting aortic wall weakening phenomena \cite{salmasi2021high}.}
\textcolor{gr}{
PCA-based statistical shape analysis could then be used to create data-driven reduced order models and derive real-time simulation results \cite{arzani2021data}.  The set of results precomputed by numerical simulation is usually based on iso-topological grids which could be obtained through the morphing approach here described. The combination of patient-specific data, machine learning and simulation results may be the key to improve growth prediction \cite{rezaeitaleshmahalleh2023computerized} and speed up decision-making in real-life medicine \cite{ paramasivam2014methodological}.}

\section{Conclusion}
This work showed that global shape features integrated with regression models could be fundamental for improving the ascending aortic aneurysm growth prediction.
An accurate growth estimate could be used to monitor the progression of the aneurysm over time and to determine the most appropriate course of treatment for a patient. By using shape features integrated with other relevant clinical information, clinicians can investigate the aneurysm shape, monitor its progression and make informed decisions about its management, such as the timing of surgery or the need for medical intervention.
Specific to this work, while PCA appears to be more suitable for exploratory data analysis and dimensionality reduction, PLS seems to more accurately predict and model the relationships between the ascending aortic shape and the growth rate.
The use of shape modes in predicting aneurysm growth is a promising approach that leverages the power of statistical shape analysis in indicating the shapes most likely to grow.
In future, a combination of shape features and numerical simulation results could be integrated to the maximum diameter threshold for the selection of patients for whom surgery is strictly required.

\section{Conflict of interest statement}
Leonardo Geronzi, Antonio Martinez, Kexin Yan and Michel Rochette were employed by Ansys France.
The other authors have no commercial, proprietary, or financial relationships that could be construed as a potential conflict of interest.

\section{Author Contributions}
L.G.: data preparation, data analysis, methodology, software, writing; A.M.: data preparation, methodology, software, review; P.H.: methodology, conceptualization;  K.Y.: data preparation; M.R.: resources, supervision; A.B.: methodology, review; S.L.:  data collection; D.M.M.: data collection; A.L.; methodology, review; O.B.: methodology; M.D.: data collection ; P.E.:data collection; J.T.: data collection, methodology; J.P.: data collection; P.P.V.: conceptualization, methodology, review;  M.E.B.: resources, review, supervision. 

All authors contributed to the article and approved the current version of the manuscript.
\section{Acknowledgments}
The work has received funding from the European Union’s Horizon 2020 research and innovation programme under the Marie Skłodowska-Curie grant agreement No 859836, MeDiTATe: “The Medical Digital Twin for Aneurysm Prevention and Treatment”.

\section*{}

\bibliographystyle{elsarticle-num}

\section{Appendix A} \label{appendixA}

\paragraph{Mesh morphing}

Mesh morphing is a technique used to modify the shape of a computational grid \cite{alexa2002recent}. Among the morphing methods available in the literature, radial basis functions (RBFs) are well known for their interpolation quality \cite {yu2011advantages}. RBFs allow to interpolate in the space a scalar function known at discrete points, called Source Points (SP). By solving a linear system of order equal to the number of SP employed \cite{biancolini2017fast}, the displacement of a mesh node in the three directions in space can be described. The approach is meshless and able to manage every element type both for surface and volume mesh.
 The interpolation function is defined as follows: 
\begin{equation}
\label{eq:RBF_eq_1}
s(\mathbf{x})=\sum_{i=1}^{N}\gamma_i \varphi \left (\left \| \mathbf{x}-\mathbf{x_{s_{i}}} \right \|\right) + h(\mathbf{x})
\end{equation}

where  $\mathbf{x}$ is a generic position in  space,  $\mathbf{x_{s_{i}}}$ the SP position, $s(\cdot)$ the function which represents a transformation $\mathbb{R}^n$ $\rightarrow$ $\mathbb{R}$,  $\varphi(\cdot)$ the radial function of order $m$, $\gamma_i$ the weight and $h(\mathbf{x})$ a polynomial term with degree $m-1$. 
The unknowns of the system, namely the polynomial coefficients and the weights $\gamma_i$ of the radial functions, are retrieved by imposing the passage of the function on the given values and an orthogonality condition on the polinomials. The linear problem can be also written in matrix form:

\begin{equation}
\label{rbf_matrix_form}
\begin{bmatrix}
\textbf{M} & \textbf{P}\\
\textbf{P}^T & \textbf{0}
\end{bmatrix}
\left\lbrace
\begin{matrix}
\boldsymbol{\gamma}\\
\boldsymbol{\beta}
\end{matrix}
\right\rbrace
=
\left\lbrace
\begin{matrix}
\textbf{g}\\
\textbf{0}
\end{matrix}
\right\rbrace
\end{equation}
in which $\textbf{M}$ is the interpolation matrix containing all the distances between RBF centres $\textbf{M}_{ij} = \varphi \left (\left \| \mathbf{x_i}-\mathbf{x_j} \right \|\right)$, $\textbf{P}$ the matrix containing the polynomial terms that has for each row $j$ the form $\textbf{P}_j = [
\begin{matrix} 1, x_{1j}, x_{2j}, ... ,x_{nj} \end{matrix} ] $ and $\textbf{g}$ the known values at SPs.
If a deformation vector field has to be fitted in 3D (space morphing), considering $h(\mathbf{x})$ as a linear polynomial made up of known $\boldsymbol{\beta}$ coefficients:

\begin{equation}
\label{eq:RBF_eq_2}
h(\mathbf{x})=\boldsymbol{\beta_1}+\boldsymbol{\beta_2}x+\boldsymbol{\beta_3}y+\boldsymbol{\beta_4}z
\end{equation}

each component of the displacement prescribed at the Source points can be interpolated as follows:
\begin{equation}
\label{eq:RBF_eq_3}
\left\{\begin{matrix}
s_x(\mathbf{x})=\sum_{i=1}^{N}\gamma_i^x\varphi\left (\left \| \mathbf{x}-\mathbf{x_{s_i}} \right \|\right)+ \beta_1^x + \beta_2^xx+\beta_3^xy+\beta_4^xz\\ 
s_y(\mathbf{x})=\sum_{i=1}^{N}\gamma_i^y\varphi\left (\left \| \mathbf{x}-\mathbf{x_{s_i}} \right \|\right)+ \beta_1^y + \beta_2^yx+\beta_3^yy+\beta_4^yz\\ 
s_z(\mathbf{x})=\sum_{i=1}^{N}\gamma_i^z\varphi\left (\left \|\mathbf{x}-\mathbf{x_{s_i}} \right \|\right)+ \beta_1^z + \beta_2^zx+\beta_3^zy+\beta_4^zz
\end{matrix}\right.
\end{equation}

When working with a mesh, the new nodal positions can be retrieved for each node as:
\begin{equation}
\label{eq:RBF_eq_4}
\boldsymbol{x_{{node}_{new}}}=\boldsymbol{x_{node}} + \begin{bmatrix}
s_x(\boldsymbol{x_{node}})\\ 
s_y(\boldsymbol{x_{node}})\\ 
s_z(\boldsymbol{x_{node}})
\end{bmatrix}
\end{equation}
where $\boldsymbol{x_{{node}_{new}}}$ and $\boldsymbol{x_{node}}$ are the vectors containing respectively the updated and original positions of the given node.

\end{document}